\documentclass{article}
\usepackage{hyperref}
\usepackage{rotating}
\usepackage{graphics}
\usepackage{amssymb}
\usepackage{lscape}
\usepackage{wrapfig}
\usepackage{color}
\usepackage{picinpar}
\usepackage[footnotesize,bf]{caption}
\usepackage{subfigure}
\usepackage{multirow}
\usepackage[latin1]{inputenc}
\usepackage{amsmath}
\usepackage[printwatermark]{xwatermark}
\usepackage{xcolor}
\usepackage{siunitx}
\usepackage{geometry}
\usepackage{comment}
\usepackage{datetime}
\usepackage{datetime}

\newcommand*\andnewline{%
        \end{tabular}
        \\[\bigskipamount]
        \begin{tabular}[t]{c}
}

\title{A primary electron beam facility at CERN}

\author{T. \r{A}kesson \textsuperscript{1},
        Y. Dutheil    \textsuperscript{2},
        L. Evans      \textsuperscript{2},         
        A. Grudiev    \textsuperscript{2},
        Y. Papaphilippou \textsuperscript{2},
        S. Stapnes    \textsuperscript{2}\\
  On behalf of the \emph{PBC-acc-e-beams}\footnote{PBC-acc-e-beams@cern.ch} \, working group \andnewline
		\textsuperscript{1}Lund University, Lund, Sweden \\
        \textsuperscript{2}CERN, Geneva, Switzerland  }

\newdate{date}{31}{05}{2018}
\date{\displaydate{date}}

\begin{document}

\maketitle

\begin{figure}[!h]
\centering
\includegraphics[width=18cm]{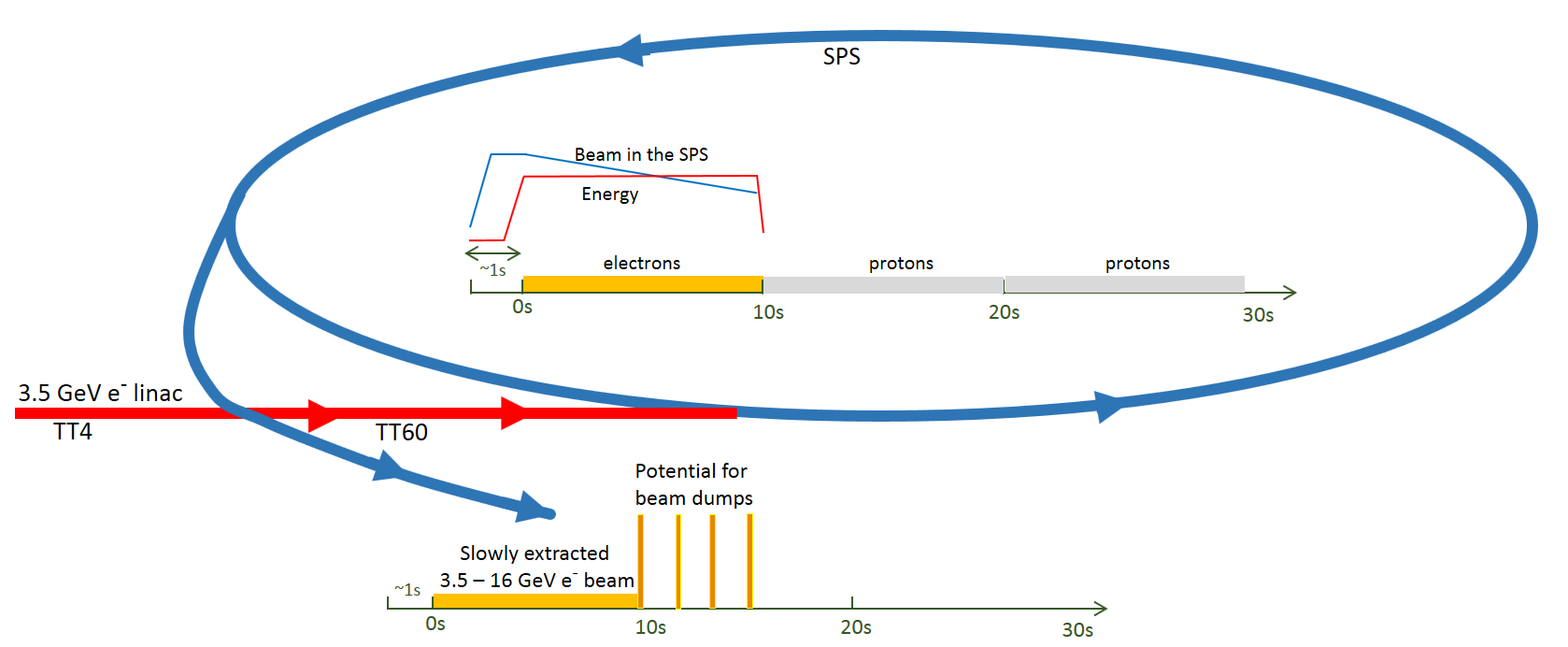}
\caption{Schematic representation of the electron beam facility at CERN with the proposed beam cycles.}
\label{fig:eSPS_layout}
\end{figure}

This document describes the concept of a primary electron beam facility at CERN, to be used for dark gauge force and light dark matter searches. The electron beam is produced in three stages: A Linac accelerates electrons from a photo-cathode up to \SI{3.5}{GeV}. This beam is injected into the Super Proton Synchrotron, SPS, and accelerated up to a maximum energy of \SI{16}{GeV}. Finally, the accelerated beam is slowly extracted to an experiment, possibly followed by a fast dump of the remaining electrons to another beamline. The beam parameters are optimized using the requirements of the Light Dark Matter eXperiment, LDMX~\cite{ldmx}, as benchmark.

\section*{Electron acceleration and extraction}

\paragraph{Electrons are produced and accelerated to \SI{3.5}{GeV}} using a high-gradient Linac that employs the technologies developed by the Compact LInear Collider (CLIC)~\cite{clic} study. 

A \SI{0.1}{GeV} S-band photo-injector produces the electron beam. Most relevant here is the laser allowing a wide range of beam time-structure to be produced at the photo-cathode RF gun.
Following the source is a \SI{3.4}{GeV} X-Band high-gradient Linac using the technology developed for the CLIC study. The design uses \SI{5.3}{m} long RF-units each capable of accelerating \SI{200}{ns} trains of 40 bunches by \SI{264}{MeV}. Each RF unit makes use of a modulator, two klystrons and a RF pulse compressor feeding power to eight copper accelerating structures. Table \ref{tab:linac_parms} summarizes the proposed beam and Linac parameters. 

\begin{table}[!ht]
\caption{Summary of possible electron beam parameters for the Linac}
\label{tab:linac_parms}
\center
\begin{tabular}{|c|c|}
\hline 
Parameter & Value \\ 
\hline 
Bunch length & ~ \SI{5}{ps} \\
Normalized transverse emittance & ~ \SI{10}{\mu m} \\
Bunch train length & \SI{200}{ns} \\
Number of bunches per train & 40 \\
Bunch charge & up to \SI{300}{pC} with 40 bunches \\
Repetition rate & \SI{100}{Hz} \\
\hline 
\end{tabular} 
\end{table}

A possible location for the \SI{70}{m} long Linac has already been identified. The Transfer Tunnel 4, TT4, building in the North-West part of the CERN Meyrin site has the required size.

\paragraph{Transfer between the Linac and the SPS} makes use of existing tunnels. The TT4 is connected to the SPS via an existing tunnel TT61 leading to the SPS Long Straight Section 6, LSS6. No civil engineering is required and the beam can be injected into the SPS using the same scheme as during the LEP era.

The SPS injection scheme features a fast bunch to bucket injection of the \SI{200}{ns} trains with a 100 Hz repetition rate directly into the SPS \SI{200}{MHz} RF-system. Preliminary studies show that the current magnetic elements used for proton extraction can be used to inject electrons at this location. Only a new fast kicker with \SI{100}{ns} rise time and \SI{200}{ns} flat-top is required but uses available space in the SPS lattice and conventional transmission line magnet technology.

\paragraph{Acceleration in the SPS} uses the same methods and components as when the SPS was used as electron injector for the LEP~\cite{SPS-LEP}. Acceleration of electrons to \SI{16}{GeV} requires \SI{10}{MV} of total RF voltage. Existing \SI{200}{MHz} \SI{1}{MV} standing-wave cavities from the LEP era can be re-installed in an available location of the SPS lattice. Twelve of those cavities are in storage and in very good conditions. Figure \ref{fig:eSPS_layout} shows that the process of injecting and accelerating in the SPS takes around \SI{1}{s}.

\paragraph{Extraction from the SPS} can be done using a resonant scheme identical to the one already used for extracting proton beam from the SPS. With an electron beam at high energy where beam dynamics is dominated by synchrotron radiation, the resonant extraction process can be driven by the intrinsic diffusion within the beam caused by quantum excitation. Third order resonant extraction is a standard techniques that has been proved capable of providing very low intensities when driven using diffusion processes~\cite{ultra_slowex}.

The proposed scheme requires installation of a new electrostatic septum, four small slow bumper magnets and two new thin magnetic septa, all in available spaces. These are standard elements already used for proton extraction.
														
\paragraph{Transfer from the SPS LSS1} uses the existing TT10 line and elements. This beamline is capable of transporting the electron beam without modification. This brings the beam back to the Meyrin site where several sites are considered for housing an experimental area at the end of the TT10 line.

\section*{Extracted beam}

The beam from the SPS can be extracted at energies anywhere from \SI{3.5}{GeV} to \SI{16}{GeV} to accommodate an LDMX (Light Dark Matter eXperiment) detector. 
\paragraph{The time structure} is constrained by the injection scheme and SPS RF system and features :
\begin{itemize}
\item a bunch spacing of \SI{5}{ns}, \SI{10}{ns} or any sub-multiple of the \SI{200}{MHz} RF bucket structure.
\item a train length of \SI{200}{ns} corresponding to up to 40 bunches per trains with a minimum of \SI{5}{ns} bunch spacing.
\item a gap of \SI{100}{ns} between trains to allow for the injection kicker rise time. 
\item a realistic spill length of \SI{10}{s} inserted into a super cycle of \SI{30}{s}, only limited by concurrent usage of the SPS for other beams.
\end{itemize}

\paragraph{The average intensity} is controlled using several methods. The slow extraction mechanism, driven by a diffusion process, allows for accurate control of the extracted intensity down to very low currents. Other collimator devices and specific optics along the extraction beam line provide further control of the beam intensity.

\paragraph{The transverse structure} is selected by careful collimation and optical choices. Collimation of the extracted beam can provide an almost flat distribution. The beam size at the target can be greatly increased and accurately controlled in both planes. For instance a \SI{2}{cm} vertical and \SI{30}{cm} horizontal beam size is feasible with conventional magnets.

\paragraph{Number of electrons on target for an LDMX-like experiment} This facility would deliver $10^{14}$ electrons per year with a \SI{20}{ns} bunch spacing and one electron extracted per bunch and per turn. Using a higher intensity of 22 electrons per bunch extracted on the target per turn and \SI{5}{ns} bunch spacing,  $10^{16}$ electrons per year are achievable.

\paragraph{Physics of an LDMX-like experiment} Explore dark matter interactions with a sensitivity that extends several orders of magnitude below the required sensitivity predicted by the thermal relic targets for representative dark matter candidates coupled to a dark photon in the sub-GeV mass range.

\paragraph{Fast beamdump} The slowly extracted beam for an LDMX-like detector would make use of less than $10^{10}$ electrons per cycles while the high gradient Linac and the SPS are capable of accelerating $10^{12}$ and possibly up to \num{5e12} electrons per cycle. A dump-type electron experiment, e.g. searching for dark photons decaying to Standard Model particles, using this intensity delivered within the SPS revolution period of \SI{23}{\mu s} can be investigated. As showed fig \ref{fig:eSPS_layout} such dump cycle could be repeated every \SI{2}{s}.

\paragraph{Nuclear Physics} Such an electron beam could also be of significant interest for the European nuclear physics community. It extends the energy range available at JLab, USA, and will allow for investigations that can advance our knowledge about hadrons and their underlying structures. Examples are the study of the momentum and spin distributions of sea quarks and in particular of gluons in the nucleons, the study of the excitation spectrum of nucleons and hyperons, and the prospect to produce mesons with exotic composition and/or exotic quantum numbers.

\paragraph{Accelerator physics} The Linac beam at \SI{3.5}{GeV} can be used by other users between slow extraction cycles. First of all, the construction and operation of the Linac is in itself a natural next step for the X-band high-gradient technology and beyond, towards a Compact LInear Collider (CLIC). Use of the electron beam is considered for plasma acceleration as both driver and probe. Positron production has been discussed and could be implemented in the facility. Positron production studies are of vital interest for any e$^+$e$^-$ machine but such a positron beam would also provide opportunities for studies of plasma acceleration of positrons. 
Several other general accelerator studies can be envisaged for this project as natural continuations of the studied carried out in the CLEAR facility today~\cite{clear}. Examples are high gradient and plasma lens studies, instrumentation and impedance studies, medical accelerator developments, component irradiation, THz acceleration, and educational activities. 
For the SPS beam one can take advantage of a small equilibrium emittance optics using the existing SPS lattice. This would allow the pursuit of final focus studies beyond the ATF2 if this becomes a priority.

\end{document}